\documentclass[a4paper,aps,prd,10pt,preprintnumbers,twocolumn,superscriptaddress,nofootinbib,amsmath,amssymb]{revtex4-1}\def\PRDonly#1{#1}\def\JHEAPonly#1{}

\usepackage{orcidlink,graphicx,cmap}
\usepackage[utf8]{inputenc}
\usepackage[T1]{fontenc}

\def\imo{i}
\newcommand{\ie}{{i.e.,}~}
\newcommand{\eg}{{e.g.,}~}

\JHEAPonly{\journal{Journal of High Energy Astrophysics}}

\begin{document}
\JHEAPonly{\sloppy\begin{frontmatter}}
\title{First few overtones probe the event horizon geometry}
\author{R. A. Konoplya \orcidlink{0000-0003-1343-9584}}
\email{roman.konoplya@gmail.com}
\PRDonly{\affiliation{Research Centre for Theoretical Physics and Astrophysics, Institute of Physics, Silesian University in Opava, Bezručovo náměstí 13, CZ-74601 Opava, Czech Republic}}
\JHEAPonly{\affiliation{organization={Research Centre for Theoretical Physics and Astrophysics, Institute of Physics, Silesian University in Opava},
            addressline={Bezručovo náměstí 13},
            city={Opava},
            postcode={CZ-74601},
            state={Moravskoslezský kraj},
            country={Czech Republic}}}

\author{A. Zhidenko \orcidlink{0000-0001-6838-3309}}
\email{olexandr.zhydenko@ufabc.edu.br}
\PRDonly{\affiliation{Centro de Matemática, Computação e Cognição (CMCC), Universidade Federal do ABC (UFABC), \\ Rua Abolição, CEP: 09210-180, Santo André, SP, Brazil}}
\JHEAPonly{\affiliation{organization={Centro de Matemática, Computação e Cognição (CMCC), Universidade Federal do ABC (UFABC)},
            addressline={Rua Abolição},
            city={Santo André},
            postcode={CEP: 09210-180},
            state={SP},
            country={Brazil}}}

\begin{abstract}
It is broadly believed that quasinormal modes cannot tell the black-hole near-horizon geometry, because usually the low-lying modes are determined by the scattering of perturbations around the peak of the effective potential.
Using the general parametrization of the black-hole spacetimes respecting the generic post-Newtonian asymptotic, we will show that tiny modifications of the Schwarzschild/Kerr geometry in a relatively small region near the event horizon lead to almost the same Schwarzschild/Kerr fundamental mode, but totally different first few overtones.
Having in mind that the first several overtones affect the quasinormal ringing at its early and intermediate stage [M. Giesler, M. Isi, M. Scheel, and S. Teukolsky, Phys. Rev. X 9, 041060 (2019)], we argue that the near-horizon geometry could in principle be studied via the first few overtones of the quasinormal spectrum, which is important because corrections to the Einstein theory must modify precisely the near-horizon geometry, keeping the known weak field regime.
\end{abstract}

\PRDonly{\maketitle}
\JHEAPonly{\end{frontmatter}}

\section{Introduction}
The proper oscillation frequencies of black holes, quasinormal modes (QNMs) \cite{Nollert:1999ji,Kokkotas:1999bd,Berti:2009kk,Konoplya:2011qq}, govern the decay of perturbations of black holes and the fundamental mode, which has the smallest decay rate, dominates at late times of the black holes' ringdown. Therefore, within the astrophysical context the fundamental mode was the one mostly studied in the literature. Despite the quasinormal frequencies do not depend on the way of perturbations, but only on the parameters of black holes, it was shown that the ringdown profile may not probe the geometry of the event horizon \cite{Cardoso:2016rao}, so that even qualitatively different compact objects, such as wormholes \cite{Cardoso:2016rao,Damour:2007ap,Konoplya:2016hmd} can mimic the black hole ringdown. This happens because the main part of the scattering from the effective potential surrounding the black hole occurs near the peak of the potential barrier and this region determines the dominant QNMs. Deformations of the geometry only near the event horizon do not alter those dominant frequencies. If the near-horizon deformation is such that an additional peak appears near the surface of the compact body, then at very late times the decay of gravitational waves are slightly modified by echoes \cite{Cardoso:2016rao}, while the fundamental mode remains again essentially unaffected. Actually the same logic is applicable to other astrophysically relevant phenomena, such as particles' motion, lensing etc., so that in addition to the weak and elusive echoes \cite{Chakraborty:2022zlq}, only the Hawking radiation tells the properties of the event horizon \cite{Damour:2007ap}.

Although it is believed that the major contribution to the signal is owing to the fundamental mode, recently it has been shown \cite{Giesler:2019uxc} (and later studied in~\cite{Oshita:2021iyn,Forteza:2021wfq,Oshita:2022pkc}) that the first several overtones must be taken into account in order to model the ringdown phase, obtained within the accurate numerical relativity simulations at the beginning of the quasinormal ringing and not only at the last stage. This finding indicated also that the actual quasinormal ringing starts earlier than expected. In \cite{Isi:2019aib} it was shown that an independent measurement of the quasinormal frequency of the first overtone leads to agreement with the no-hair hypothesis at the $20\%$ level. When the detector sensitivity and the observable population of black-hole mergers increase, one can expect that overtones will provide more efficient tests of the strong gravity regime \cite{Isi:2019aib}.

Once the overtones are taken into account, the modelling and linear profile of the ringdown are in full concordance and allow to extract angular mass and momentum of the black hole. The posterior analysis of LIGO's observational data from the GW150914 gravitational-wave signal \cite{LIGOScientific:2016dsl} provides evidence for the presence of the first overtone rather than noise \cite{Isi:2021iql,Crisostomi:2023tle}. Therefore, detection of the higher overtones is expected in the near future as the signal-to-noise ratio improves with the next-generation gravitational-wave detectors, such as the Einstein Telescope \cite{Punturo:2010zz}, Cosmic Explorer \cite{Dwyer:2014fpa}, and the space-based interferometer LISA \cite{LISA:2017pwj}. Although the current sensitivity of the LIGO/VIRGO detectors does not yet allow for the unambiguous detection of the overtones in the gravitational-wave signals \cite{Capano:2021etf,Cotesta:2022pci} and the non-linear corrections must be taken into account when analyzing the higher-overtones contributions \cite{Sberna:2021eui,Cheung:2022rbm,Mitman:2022qdl}, there are indications that the overtones are significantly excited in some events and can be detected by LISA at the early ringdown phase \cite{Oshita:2022yry}.

Having in mind the above motivation, we will consider the general parametrized spherically and axially symmetric black holes, respecting the experimentally allowed post-Newtonian behavior \cite{EventHorizonTelescope:2021dqv}. There are two sets of parameters: experimentally constrained ones, determining the weak field regime at a distance from the black hole and the ones fixed by the near-horizon strong gravity region. We show that deformations of the near-horizon parameters induce outburst of overtones, while the fundamental mode changes insignificantly. This observation is important, because essentially non-Einsteinian behavior is expected exactly near the event horizon, either due to quantum corrections or alternative theories of gravity (see \eg\cite{Konoplya:2022hll}), including the brane-world models. Since the fundamental mode should not be affected strongly by such near-horizon corrections, one can expect that the dominant nonlinear effects due to this mode \cite{Sberna:2021eui,Cheung:2022rbm,Mitman:2022qdl} do not lead to the significant deviations from the Kerr ringdown profile as well.

This way, the first few overtones, which potentially could be extracted from the beginning of the time-domain profile of quasinormal ringing could probe the near-horizon geometry of a black hole and, for example, distinguish a black-hole mimicker. The latter might be more profitable way than echoes, which occur at very late times when the amplitude of the signal is strongly damped.

Here we would like to provide a brief clarification on terminology. When discussing the alterations to the geometry of a black hole relative to its Schwarzschild/Kerr limit, induced by alternative theories or other factors, the term ``perturbations'' is occasionally employed. In fact, what is typically referred to as ``perturbations'' in those papers \cite{Jaramillo:2020tuu,Jaramillo:2021tmt,Cheung:2021bol} is merely the addition of a small function of the radial coordinate to the initial black hole metric, such as the Schwarzschild metric. This function is independent of time. Specifically:
\begin{equation}
\begin{array}{lr}
g_{\mu \nu}(r, \theta, \phi) \rightarrow g_{\mu \nu}(r, \theta, \phi) + \delta g_{\mu \nu}(r) & \qquad \mbox{-- deformation.}
\end{array}
\end{equation}
The perturbation of any given static or stationary black hole spacetime, however, must be time-dependent and is responsible for dynamical degrees of freedom:
\begin{equation}
\begin{array}{lr}
g_{\mu \nu}(r, \theta, \phi) \rightarrow g_{\mu \nu}(r, \theta, \phi) + \delta g_{\mu \nu}(t, r, \theta, \phi) & \mbox{-- perturbation}.
\end{array}
\end{equation}
The mere smallness of the added term is not sufficient to call it a ``perturbation''. This addition does not correspond to any dynamical degree of freedom; rather, it simply modifies the background metric geometry in a static manner. When considering a stationary metric, such as Kerr, as the ``non-perturbed background'', the added term remains stationary and time-independent.

Therefore, we find this term misleading as it may be mistaken for perturbations of an already deformed stationary geometry of a black hole. Similarly, the set of perturbations for these deformed non-Schwarzschild/non-Kerr black holes is sometimes referred to as a ``quasi-spectrum'', despite it being the typical spectrum of quasinormal modes for a specific black hole geometry. Hence, we opt to refrain from using the terms ``perturbations'' and ``quasi-spectrum'', instead preferring ``deformation'' and ``spectrum of the deformed black hole'', respectively. The same logic applies to the term ``(in)stability'' of the frequencies. This refers to the fact that the frequencies of the deformed black hole deviate from their Schwarzschild values at an increasing rate while the deformation remains small. Such ``(in)stability'' can be confused with the concept of dynamical instability caused by small time-dependent perturbations, where certain modes grow unboundedly over time, leading to a situation where the entire configuration becomes unsustainable.

\section{Basic equations}
\subsection{The general parametrized black-hole metric}

The metric of a spherically symmetric black hole can be written in the following general form,
\begin{equation}
ds^2=-N^2(r)dt^2+\frac{B^2(r)}{N^2(r)}dr^2+r^2 (d\theta^2+\sin^2\theta d\phi^2),\label{metric}
\end{equation}
where $r_0$ is the event horizon, so that $N(r_0)=0$.
Following \cite{Rezzolla:2014mua}, we will use the new dimensionless variable
$$x \equiv 1 - \frac{r_0}{r},$$ so that $x=0$ corresponds to the event horizon, while $x=1$ corresponds to spatial infinity. We rewrite the metric function $N$ via the expression $N^2=x A(x)$, where $A(x)>0$ for \mbox{$0\leq x\leq1$}. Using the new parameters $\epsilon$, $a_0$, and $b_0$, the functions $A$ and $B$ can be written as
\begin{eqnarray}\nonumber
A(x)&=&1-\epsilon (1-x)+(a_0-\epsilon)(1-x)^2+{\tilde A}(x)(1-x)^3\,,
\\
B(x)&=&1+b_0(1-x)+{\tilde B}(x)(1-x)^2\,.\label{ABexp}
\end{eqnarray}
Here the coefficient $\epsilon$ measures the deviation of $r_0$ from the Schwarzschild radius $2 M$:
$$\epsilon = \frac{2 M-r_0}{r_0}.$$
The coefficients $a_0$ and $b_0$ can be considered as combinations of the post-Newtonian (PN) parameters,
$$
a_0=\frac{1}{2}(\beta-\gamma)(1+\epsilon)^2,
\qquad
b_0=\frac{1}{2}(\gamma-1)(1+\epsilon).
$$
Current observational constraints on the PN parameters imply \mbox{$a_0 \sim b_0 \sim 10^{-4}$}, so that we can safely neglect them.

The functions ${\tilde A}$ and ${\tilde B}$ are introduced through infinite continued fraction in order to describe the metric near the horizon (\ie for $x \simeq 0$),
\begin{equation}\label{ABdef}
{\tilde A}(x)=\dfrac{a_1}{1+\dfrac{a_2x}{1+\dfrac{a_3x}{1+\ldots}}}, \quad
{\tilde B}(x)=\dfrac{b_1}{1+\dfrac{b_2x}{1+\dfrac{b_3x}{1+\ldots}}},
\end{equation}
where $a_1, a_2,\ldots$ and $b_1, b_2,\ldots$ are dimensionless constants to be constrained from observations of phenomena which are localized near the event horizon. At the horizon only the first term in each of the continued fractions survives,
$$
{\tilde A}(0)={a_1},\qquad
{\tilde B}(0)={b_1},
$$
which implies that near the horizon only the lower-order terms of the expansions are essential. For instance, the Hawking temperature depends on the coefficients of the lowest order only \cite{Konoplya:2021qll},
\begin{eqnarray}
    T_H\equiv\frac{\kappa_g}{2\pi}=\frac{A(0)}{4\pi B(0)}=\frac{1-2\epsilon+a_0+a_1}{4\pi r_0(1+b_0+b_1)}.
\end{eqnarray}

When all the coefficients $a_i$, $b_i$ and $\epsilon$ vanish, we have the Schwarzschild black hole, so that one can consider the above parametrization as a general deformation of the Schwarzschild geometry. Then, we would like to understand which kind of deformations are responsible for the outburst of overtones. For this purpose, following \cite{Konoplya:2020hyk}, we will distinguish the Schwarzschild-like \textit{nonmoderate} black holes, whose metric functions are close to the Schwarzschild one everywhere except for a relatively small region near the event horizon, in which it is strongly different.
The word \textit{relatively} means here that the deformation must decay and be negligibly small near the peak of the potential barrier. It is believed that this kind of black holes can be Schwarzschild mimickers, so that their ringdown profile and shadows are practically indistinguishable from those for the Schwarzschild ones \cite{Konoplya:2020hyk,Konoplya:2021slg}. {\it Moderate} black holes are characterized by relatively slow change of the metric functions in the near horizon zone.

\subsection{Calculation of the quasinormal modes}
After the separation of variables in the general covariant equations for the scalar and electromagnetic perturbations they can be reduced to the Schrödinger-like form (see, \eg \cite{Konoplya:2011qq}),
\begin{equation}\label{wave-equation}
\dfrac{\partial^2 \Psi}{\partial t^2}-\dfrac{\partial^2 \Psi}{\partial r_*^2}+V(r)\Psi=0,
\end{equation}
where the ``tortoise coordinate'' $r_*$ is defined by the relation $$dr_*=\dfrac{B(r)dr}{N^{2}(r)}.$$
The effective potentials for the scalar and electromagnetic fields are
$$V(r)=N^2(r)\frac{\ell(\ell+1)}{r^2} + \frac{1-s}{2r}\frac{d}{dr}\frac{N^4(r)}{B^2(r)},$$
where $\ell=1, 2, \ldots$ are the multipole numbers and $s=0$ ($s=1$) corresponds to the scalar (electromagnetic) field, respectively. The effective potential for the electromagnetic field has the form of the positive definite potential barrier, while this is not always so for a scalar field.

Quasinormal modes $\omega_{n}$ are frequencies corresponding to solutions of the master wave equation (\ref{wave-equation}) with the requirement of the purely outgoing waves at infinity and at the event horizon, $\Psi \propto e^{-\imo \omega t \pm \imo \omega r_*}, r_* \to \pm \infty$.
In order to find QNMs we will use two methods: the time-domain integration and the Frobenius method.

In the time domain, we integrate the wavelike equation (\ref{wave-equation}) in terms of the light-cone variables $u=t-r_*$ and $v=t+r_*$,
using the discretization scheme of \cite{Gundlach:1993tp} and, further, extracting QNMs with the Prony method \cite{Prony}.

In the frequency domain, after separating the time and radial coordinate, $\Psi(t,r)=e^{-\imo \omega t}R(r)$, we use the Frobenius method~\cite{Leaver:1985ax}. Namely, we express the function $R$, written with respect to the compact coordinate $x$, as a product of the factor, which diverges at the singular points $x=0$ and $x=1$ satisfying the quasinormal boundary conditions, and the Frobenius series expansion
\begin{equation}
R(x)=x^{-\imo\omega/2\kappa_g}e^{\imo\omega r}r^{\lambda}\sum_{m=0}^\infty c_m x^m,
\end{equation}
where $\kappa_g>0$ and $\lambda$ are determined by substituting the series into (\ref{wave-equation}) and expanding, respectively, at $x=0$ and $x=1$.
By expanding the wavelike equation at $x=0$, we find the recurrence relation for the coefficients $c_m$, which can be numerically reduced to the three-terms relation via Gaussian eliminations. Finally, we obtain an equation with the infinite continued fraction with respect to $\omega$. In order to calculate the infinite continued fraction we use the Nollert improvement~\cite{Nollert:1993zz}, which was generalized in~\cite{Zhidenko:2006rs} for an arbitrary number of terms in the recurrence relation (see Sec.~3.4~of~\cite{Zhidenko:2009zx}). When the singular points of the wavelike equation appear within the unit circle $|x|<1$, we employ a sequence of positive real midpoints as described in \cite{Rostworowski:2006bp}.

\subsection{Gravitational perturbations of rotating black holes}

The general parametrization of spherically symmetric black holes \cite{Rezzolla:2014mua} was extended to the axially-symmetric case in \cite{Konoplya:2016jvv}. Here, for illustration, we will consider the near-horizon deformations of the Kerr metric which fall into a particular, more symmetric class of this general parametrization. The parametrized metric includes deformations in the radial and axial directions and has the general form given by eqs. (1, 22, 67) of \cite{Konoplya:2018arm}.

The wave-like equation has the form \cite{Kanti:2006ua},
\begin{eqnarray}\label{radialKerr}
&& \Delta(r)^{s}(r)\,\frac{d}{dr}\,\left(\Delta(r)^{1-s}\,\frac{d R}{dr}\,\right)
\\
&& + \left(\frac{K^2(r)-isK(r) \Delta'(r)}{\Delta(r)} + 4i s\,\omega\,r
- \lambda\right)\,R(r)=0\,,\nonumber
\label{radial}
\end{eqnarray}
where
\begin{eqnarray}\nonumber
\Delta(r) &\equiv& (r^2A(1-r_0/r)+a^2)(1-r_0/r),
\\\nonumber
K(r)&\equiv&(r^2+a^2)\omega-am,
\end{eqnarray}
and $\lambda$ is the separation constant. Here for the gravitational perturbations we take \mbox{$s=-2$}. Then we consider ad hoc deformation of the wave-like equation for gravitational perturbations of Kerr black hole via deformations implemented in $A(x)$ (\ref{ABexp}).

\section{How to prepare deformations solely in the near-horizon zone}
\subsection{Deformation of the black hole}

In this paper we will study the nonmoderate Schwarzschild-like black holes, which have the Schwarzschild size-to-mass ratio and satisfy the post-Newtonian constraints $\epsilon=a_0=b_0$. This choice implies that the black-hole geometry is almost Schwarzschildian everywhere starting from some distance from the black hole, which includes not only the asymptotic post-Newtonian region, but also the region around the maximum of the wave equation effective potential, photon sphere and the innermost stable circular orbit, that is, the region responsible for the dominant characteristics of various astrophysical radiation phenomena. The near-horizon geometry of such nonmoderate black holes differs significantly from the Schwarzschild ones: for instance, the nonmoderate black hole can have different Hawking temperature and/or large values of higher derivatives of the metric function, corresponding to large values of the near-horizon coefficients $a_1, a_2,\ldots$ and $b_1, b_2,\ldots$. Overall, here we will compare the Schwarzschild black hole with the following four cases:
\begin{enumerate}
\item Nonmoderate Schwarzschild-like black hole with the Schwarzschild values of the Hawking temperature and radius of the event horizon (black hole 1).
\item Nonmoderate Schwarzschild-like black hole with considerably different (from the Schwarzschild one) Hawking temperature, but the same radius of the event horizon (black hole 2).
\item Moderate black hole with a slightly different radius, but the same mass and post-Newtonian behavior (black hole 3).
\end{enumerate}
Thus, we study the moderate and nonmoderate near-horizon deformations of the Schwarzschild geometry.

\subsection{Deformations of the wave-equation}
In addition to the deformations of the wavelike equation due to deviation of the black-hole geometry from the General Relativity solution (Schwarzschild or Kerr geometry), we will consider the deformations due to some external (with respect to the black hole) source. These sources can lead to deformations in the near-horizon region (\eg due to the strong-field quantum effects) or at a distance (\eg due to accreting matter).
The second situation is the most interesting, because quantum corrections in the near horizon zone in the form of bath of quantum matter fields under supposition of purely Einsteinian gravity would have incremental influence upon the metric, being localized in the tiny layer of order of Planck length and with negligible energy density. This physically non-motivated configuration was considered in \cite{Cardoso:2024mrw}. Here, we have in mind quantum corrections which, first of all, would modify the gravitational sector. Thus, we either mean astrophysical environment at a distance from the black hole or modification of gravitational sector in the near-horizon zone.

We shall consider the simplest kind of the above deformations by adding an augmentation to the effective potential in a form of a (smaller) peak,
\begin{equation}
V\mapsto V+\delta V,
\end{equation}
where
\begin{equation}\label{PT-augmentation}
\delta V = \frac{\delta}{r_0^2}\cosh^{-2}\frac{r_*-r_p}{r_s},
\end{equation}
where $\delta$, $r_p$, and $r_s$ are constants, defining, respectively, size, position, and wideness of the augmentation.

The Pöschl-Teller-like augmentation (\ref{PT-augmentation}) not only add a localized deformation in the vicinity of $r_*=r_p$, but also changes the singular point of the wavelike differential equation (\ref{wave-equation}), which corresponds to the event horizon $r=r_0$ ($r_*=-\infty$). This is because the hyperbolic cosine has irregular singular point at $r_*=\infty$. That is why, in order to model the deformation which does not affect the horizon, we consider another augmentation in the form of a rational function of $r$, which approaches zero at the horizon as $\delta V\propto(r-r_0)^h$ and at infinity as $\delta V\propto r^{-a}$, having the maximum value $\delta V_{max}=\delta/r_0^2$ at $r=r_m>r_0$,
\begin{equation}\label{augm}
\delta V=\frac{\delta}{r_0^2}\left(\frac{1-r_0/r}{1-r_0/r_m}\right)^h\left(1+\frac{h}{a}\times\frac{r/r_m-1}{r_m/r_0-1}\right)^{-a}.
\end{equation}

\section{Overtones' outburst}

\begin{table*}
\resizebox{\textwidth}{!}{
\begin{tabular}{|c@{\hspace{.3em}}|@{\hspace{.3em}}c@{\hspace{.5em}}c@{\hspace{.5em}}c@{\hspace{.3em}}|@{\hspace{.3em}}c@{\hspace{.5em}}c@{\hspace{.5em}}c@{\hspace{.3em}}|@{\hspace{.3em}}c@{\hspace{.5em}}c|}
\hline
\hline
Schwarzschild & black hole 1 & $D_{Re}$ & $D_{Im}$ & black hole 2 & $D_{Re}$ & $D_{Im}$ & black hole 3 & $D$\\
\hline
\multicolumn{9}{|c|}{$s=0$, $\ell=1$}\\
\hline
$0.585872 - 0.195320\imo$ & $0.587258 - 0.197245\imo$ & $0.24\%$ & $0.99\%$ & $0.587765 - 0.193327\imo$ & $0.32\%$ & $1.02\%$ & $0.583061 - 0.195929\imo$ & $0.466\%$\\
$0.528897 - 0.612515\imo$ & $0.526080 - 0.617827\imo$ & $0.53\%$ & $0.87\%$ & $0.561322 - 0.568479\imo$ & $6.13\%$ & $7.19\%$ & $0.524958 - 0.614911\imo$ & $0.570\%$\\
$0.459079 - 1.080267\imo$ & $0.447096 - 1.087917\imo$ & $2.61\%$ & $0.71\%$ & $0.679860 - 0.931860\imo$ & $48.1\%$ & $13.7\%$ & $0.453920 - 1.085471\imo$ & $0.624\%$\\
$0.406517 - 1.576596\imo$ & $0.379277 - 1.585328\imo$ & $6.70\%$ & $0.55\%$ & $0.898879 - 1.342045\imo$ & $121 \%$ & $14.9\%$ & $0.400556 - 1.584962\imo$ & $0.631\%$\\
$0.370218 - 2.081524\imo$ & $0.319572 - 2.090056\imo$ & $13.7\%$ & $0.41\%$ & $1.144760 - 1.754668\imo$ & $209 \%$ & $15.7\%$ & $0.363661 - 2.093076\imo$ & $0.628\%$\\
$0.344154 - 2.588239\imo$ & $0.256746 - 2.595787\imo$ & $25.4\%$ & $0.29\%$ & $1.404472 - 2.162571\imo$ & $308 \%$ & $16.4\%$ & $0.337075 - 2.602974\imo$ & $0.626\%$\\
$0.324452 - 3.094880\imo$ & $0.164957 - 3.102742\imo$ & $49.2\%$ & $0.25\%$ & $1.671835 - 2.566564\imo$ & $415 \%$ & $17.1\%$ & $0.316883 - 3.112805\imo$ & $0.625\%$\\
\hline
\multicolumn{9}{|c|}{$s=1$, $\ell=1$}\\
\hline
$0.496527 - 0.184975\imo$ & $0.497286 - 0.186510\imo$ & $0.15\%$ & $0.83\%$ & $0.499888 - 0.184026\imo$ & $0.68\%$ & $0.51\%$ & $0.493495 - 0.185416\imo$ & $0.578\%$\\
$0.429031 - 0.587335\imo$ & $0.425875 - 0.591467\imo$ & $0.74\%$ & $0.70\%$ & $0.477337 - 0.570306\imo$ & $11.3\%$ & $2.90\%$ & $0.424652 - 0.589368\imo$ & $0.664\%$\\
$0.349547 - 1.050375\imo$ & $0.337877 - 1.056161\imo$ & $3.34\%$ & $0.55\%$ & $0.579548 - 0.976294\imo$ & $65.8\%$ & $7.05\%$ & $0.343812 - 1.055308\imo$ & $0.683\%$\\
$0.292353 - 1.543818\imo$ & $0.267031 - 1.550075\imo$ & $8.66\%$ & $0.41\%$ & $0.783696 - 1.404047\imo$ & $168 \%$ & $9.05\%$ & $0.285850 - 1.552058\imo$ & $0.668\%$\\
$0.253108 - 2.045101\imo$ & $0.208316 - 2.050006\imo$ & $17.7\%$ & $0.24\%$ & $1.019914 - 1.827973\imo$ & $302 \%$ & $10.6\%$ & $0.246129 - 2.056707\imo$ & $0.657\%$\\
$0.224506 - 2.547851\imo$ & $0.153282 - 2.548313\imo$ & $31.7\%$ & $0.02\%$ & $1.270965 - 2.244820\imo$ & $466 \%$ & $11.9\%$ & $0.217204 - 2.562872\imo$ & $0.653\%$\\
$0.202429 - 3.050533\imo$ & $0.096081 - 3.040733\imo$ & $52.5\%$ & $0.32\%$ & $1.530215 - 2.656275\imo$ & $656 \%$ & $12.9\%$ & $0.194916 - 3.069024\imo$ & $0.653\%$\\
\hline
\multicolumn{9}{|c|}{$s=0$, $\ell=2$}\\
\hline
$0.967288 - 0.193518\imo$ & $0.970301 - 0.195678\imo$ & $0.31\%$ & $1.12\%$ & $0.968930 - 0.193534\imo$ & $0.17\%$ & $0.01\%$ & $0.962830 - 0.194098\imo$ & $0.456\%$\\
$0.927701 - 0.591208\imo$ & $0.927929 - 0.597500\imo$ & $0.03\%$ & $1.06\%$ & $0.890084 - 0.593171\imo$ & $4.06\%$ & $0.33\%$ & $0.922453 - 0.593168\imo$ & $0.509\%$\\
$0.861088 - 1.017117\imo$ & $0.855454 - 1.027058\imo$ & $0.65\%$ & $0.98\%$ & $0.895286 - 0.760648\imo$ & $3.97\%$ & $25.2\%$ & $0.854530 - 1.021068\imo$ & $0.575\%$\\
$0.787726 - 1.476193\imo$ & $0.772643 - 1.489215\imo$ & $1.91\%$ & $0.88\%$ & $1.115236 - 1.148286\imo$ & $41.6\%$ & $22.2\%$ & $0.779838 - 1.482817\imo$ & $0.616\%$\\
$0.722598 - 1.959843\imo$ & $0.693764 - 1.975524\imo$ & $3.99\%$ & $0.80\%$ & $1.320778 - 1.558055\imo$ & $82.8\%$ & $20.5\%$ & $0.713655 - 1.969520\imo$ & $0.631\%$\\
$0.669799 - 2.456822\imo$ & $0.621746 - 2.475224\imo$ & $7.17\%$ & $0.75\%$ & $1.550726 - 1.970153\imo$ & $131 \%$ & $19.8\%$ & $0.660043 - 2.469663\imo$ & $0.633\%$\\
$0.627772 - 2.959909\imo$ & $0.552750 - 2.982487\imo$ & $12.0\%$ & $0.76\%$ & $1.796557 - 2.380874\imo$ & $186 \%$ & $19.6\%$ & $0.617342 - 2.975931\imo$ & $0.632\%$\\
\hline
\multicolumn{9}{|c|}{$s=1$, $\ell=2$}\\
\hline
$0.915191 - 0.190009\imo$ & $0.917791 - 0.192048\imo$ & $0.28\%$ & $1.07\%$ & $0.916726 - 0.189735\imo$ & $0.16\%$ & $0.14\%$ & $0.910635 - 0.190547\imo$ & $0.491\%$\\
$0.873085 - 0.581420\imo$ & $0.872888 - 0.587320\imo$ & $0.02\%$ & $1.02\%$ & $0.847080 - 0.565277\imo$ & $2.98\%$ & $2.78\%$ & $0.867671 - 0.583268\imo$ & $0.545\%$\\
$0.802373 - 1.003175\imo$ & $0.796272 - 1.012369\imo$ & $0.76\%$ & $0.92\%$ & $0.866621 - 0.795077\imo$ & $8.01\%$ & $20.7\%$ & $0.795527 - 1.006997\imo$ & $0.610\%$\\
$0.725190 - 1.460397\imo$ & $0.709557 - 1.472175\imo$ & $2.16\%$ & $0.81\%$ & $1.060272 - 1.178507\imo$ & $46.2\%$ & $19.3\%$ & $0.716892 - 1.466942\imo$ & $0.648\%$\\
$0.657473 - 1.943219\imo$ & $0.628022 - 1.956894\imo$ & $4.48\%$ & $0.70\%$ & $1.261355 - 1.595636\imo$ & $91.8\%$ & $17.9\%$ & $0.648021 - 1.952893\imo$ & $0.659\%$\\
$0.602986 - 2.439430\imo$ & $0.554449 - 2.454492\imo$ & $8.05\%$ & $0.62\%$ & $1.487023 - 2.014416\imo$ & $146 \%$ & $17.4\%$ & $0.592630 - 2.452356\imo$ & $0.659\%$\\
$0.559690 - 2.941586\imo$ & $0.485002 - 2.958068\imo$ & $13.3\%$ & $0.56\%$ & $1.728792 - 2.430765\imo$ & $208 \%$ & $17.4\%$ & $0.548569 - 2.957790\imo$ & $0.656\%$\\
\hline
\hline
\end{tabular}
}
\caption{The dominant mode and first six overtones calculated using the Frobenius method in the units $r_0=2M=1$ ($\epsilon=0$) for the Schwarzschild black hole ($a_1=b_1=0$, $T_H^{(0)}=1/4\pi$), the black hole 1: $a_1=0.0001$, $a_2=-1000$, $a_3=1001$, $a_4=0$, $b_1=0$ ($T_H=1.0001T_H^{(0)}$), the black hole 2: $a_1=0.5$, $a_2=100$, $a_3=0$, $b_1=0$ ($T_H=1.5T_H^{(0)}$), and the black hole 3 with $\epsilon=-0.01$ ($2M=1$, $a_1=b_1=0$), and the relative differences (in per cents) of the corresponding modes from the Schwarzschild ones.}\label{tabl:overtones}
\end{table*}

\begin{figure*}
\resizebox{\linewidth}{!}{\includegraphics{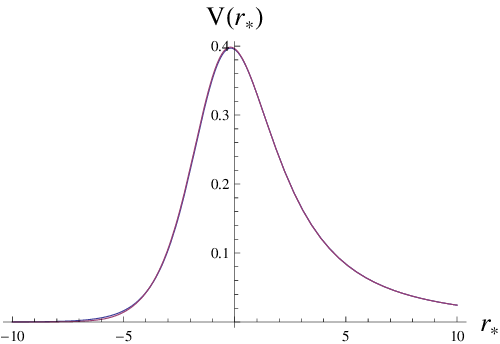}\includegraphics{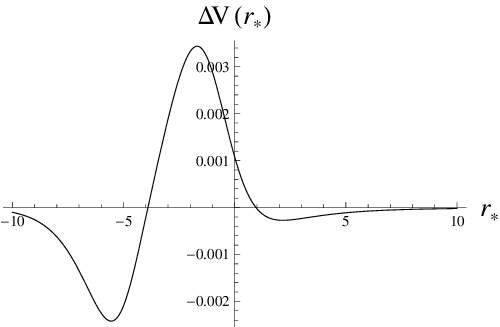}}
\caption{Left panel: Effective potentials ($s=0$, $\ell=1$, $r_0=2M=1$) for the black hole 2 (blue) and Schwarzschild black hole (red) for comparison. Right panel: difference between the potentials.}\label{fig:scalarpot}
\end{figure*}

In table~\ref{tabl:overtones} we compare accurate values of seven dominant modes of the Schwarzschild black hole and three modified black holes: two nonmoderate black holes of the same size and mass (black holes 1 and 2), which are characterized by large values of the coefficient $a_2$, while the coefficient $a_1$ can be small (then the Hawking temperature is close to the Schwarzschild one) or large (when the Hawking temperature differs a lot).
At the same time, by requiring the parameter $a_2$ to be large, the effective potential will acquire deformation near the event horizon, but remain almost the same at a distance from the black hole. In both cases the real part of the higher overtones differ significantly from the Schwarzschild spectrum. It is essential that the latter case corresponds to a very small deformation of the effective potential (fig.~\ref{fig:scalarpot}), so that even tiny deformations near the event horizon are sufficient for the outburst of overtones. When $a_1$ is not small, the difference is larger, and the spacing of the imaginary part also differs considerably, leading to quite a large difference already in the second overtone. The fundamental mode and overtones of the moderate black hole 3 change softly, not showing high sensitivity of overtones relatively the lowest mode. There remains the other (a trivial in a sense) case which we did not consider: if the asymptotic parameter $\epsilon$ differs a lot from zero, which leads to large deviations of both the fundamental mode and overtones from their Schwarzschild values.

From the data presented in table \ref{tabl:overtones}, we can see that the black hole whose Hawking temperature differs by $1.5$ times from the Schwarzschild's one, the effect of the deviation of overtones is strong already at $n=2$ case. As pointed out in \cite{Cardoso:2024mrw}, the second overtone of the black hole 2 for $\ell=2$ was missed in the first version of our paper. This mode has almost the same real part as $n=1$ while imaginary part is significantly differs both from $n=1$ and $n=2$ Schwarzschild modes. This case shows that not only the real part of $\omega$, but also the damping rate may experience an outburst.

\begin{table}
\begin{tabular}{c@{\hspace{0.5em}}c@{\hspace{1em}}c@{\hspace{0.3em}}c@{\hspace{0.3em}}c}
\hline
\hline
$n$ & Kerr & modified Kerr & $D_{Re}$ & $D_{Im}$\\
\hline
$0$ & $0.586017 - 0.075630\imo$ & $0.590393 - 0.075233\imo$ & $0.75\%$ & $0.52\%$ \\
$1$ & $0.577922 - 0.228149\imo$ & $0.578766 - 0.226295\imo$ & $0.15\%$ & $0.81\%$ \\
$2$ & $0.562240 - 0.383895\imo$ & $0.548651 - 0.391231\imo$ & $2.42\%$ & $1.91\%$ \\
$3$ & $0.538956 - 0.542888\imo$ & $0.488262 - 0.616589\imo$ & $9.41\%$ & $13.6\%$ \\
$4$ & $0.506263 - 0.697962\imo$ & $0.597420 - 0.814367\imo$ & $18.0\%$ & $16.7\%$ \\
$5$ & $0.486283 - 0.830803\imo$ & $0.709014 - 0.975480\imo$ & $45.8\%$ & $17.4\%$ \\
$6$ & $0.499165 - 0.983084\imo$ & $0.829156 - 1.138613\imo$ & $66.1\%$ & $15.8\%$ \\
$7$ & $0.506652 - 1.156708\imo$ & $0.951035 - 1.299293\imo$ & $87.7\%$ & $12.3\%$ \\
\hline
\hline
\end{tabular}
\caption{Dominant modes of gravitational perturbations ($\ell=m=2$) for the Kerr black hole $a=0.8$, $M=1$, $r_0=1.6$ ($\epsilon=0.25$) calculated in~\cite{Berti:2005ys} compared to the modified Kerr ($a_1=0.5$, $a_2=100$, $a_3=0$) and their relative differences (in per cents).}\label{tabl:grav}
\end{table}

In table~\ref{tabl:grav} one can see that the same high sensitivity of overtones takes place for the near-horizon deformations of the equation for the gravitational perturbations of rotating black holes.

\PRDonly{
\begin{figure}
\resizebox{\linewidth}{!}{\includegraphics{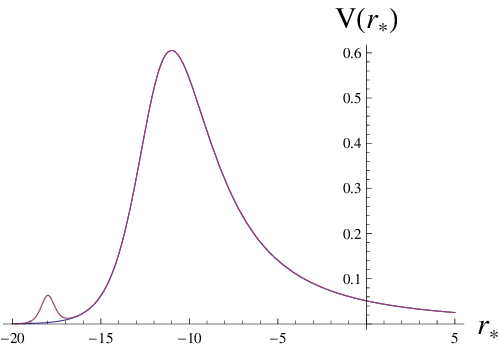}}\\
\resizebox{\linewidth}{!}{\includegraphics{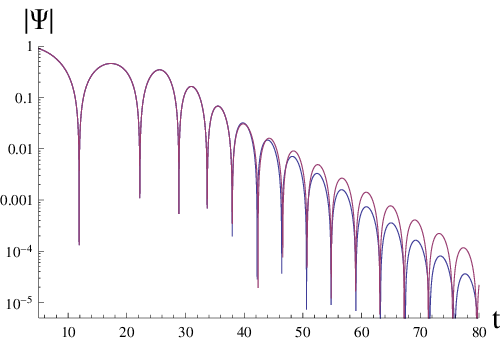}}\\
\resizebox{\linewidth}{!}{\includegraphics{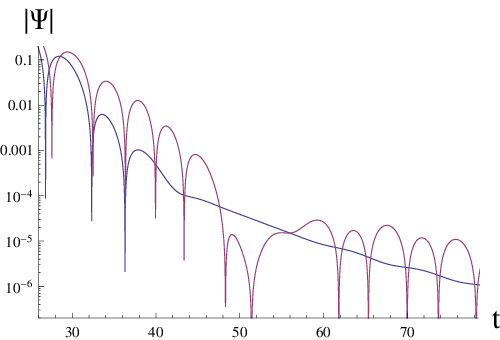}}
\caption{Upper panel: The effective potential at \mbox{$r_*=0$} corresponding to \mbox{$r=10$} for the \mbox{$\ell=2$} axial gravitational perturbations of the Schwarzschild \mbox{($r_0=2M=1$)} black hole (blue) and the potential deformed by a Pöschl-Teller-like augmentation \mbox{$\delta V=0.06/\cosh^2(2r_*+36)$} (red). Middle panel: the corresponding time-domain profiles. Lower panel: the time-domain profiles without the dominant-mode contribution, calculated with the Prony method, \mbox{$\omega_0\approx0.7473-0.1779\imo$} for the Schwarzschild black hole and \mbox{$\omega_0\approx0.7622 - 0.1491\imo$} for the deformed potential. The amplitude, corresponding to the first overtone, is about four orders larger than the amplitude of the echo, appearing at $t\simeq60$ (red). For an illustration, the augmentation is chosen to be large in order to compare the overtones effect with the echoes at earlier times. }\label{fig:gravpotdef}
\end{figure}
}
\JHEAPonly{
\begin{figure*}
\parbox{.33\linewidth}{\resizebox{\linewidth}{!}{\includegraphics{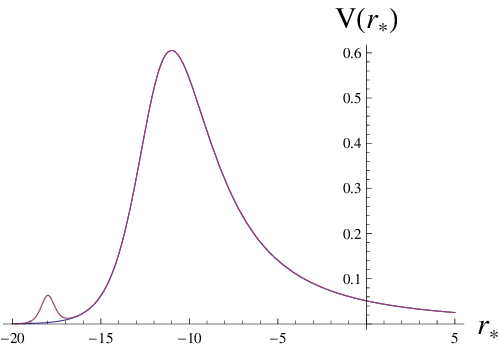}}\\\resizebox{\linewidth}{!}{\includegraphics{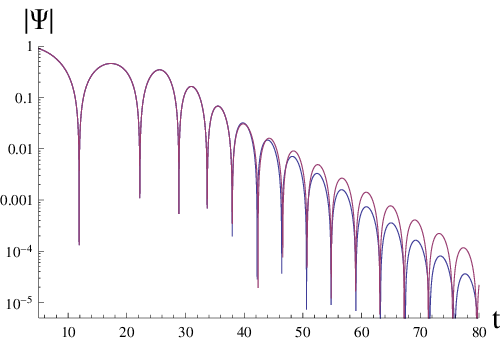}}}
\parbox{.65\linewidth}{\resizebox{\linewidth}{!}{\includegraphics{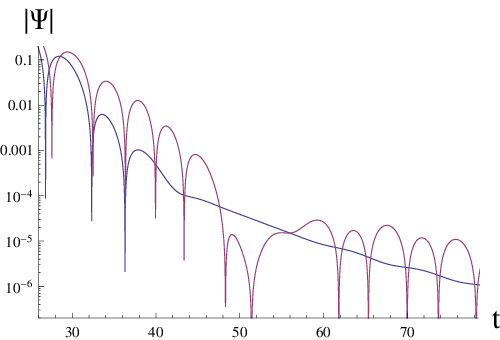}}}
\caption{Upper left panel: The effective potential at \mbox{$r_*=0$} corresponding to \mbox{$r=10$} for the \mbox{$\ell=2$} axial gravitational perturbations of the Schwarzschild \mbox{($r_0=2M=1$)} black hole (blue) and the potential deformed by a Pöschl-Teller-like augmentation \mbox{$\delta V=0.06/\cosh^2(2r_*+36)$} (red). Lower left panel: the corresponding time-domain profiles. Right panel: the time-domain profiles without the dominant-mode contribution, calculated with the Prony method, \mbox{$\omega_0\approx0.7473-0.1779\imo$} for the Schwarzschild black hole and \mbox{$\omega_0\approx0.7622 - 0.1491\imo$} for the deformed potential. The amplitude, corresponding to the first overtone, is about four orders larger than the amplitude of the echo, appearing at $t\simeq60$ (red). For an illustration, the augmentation is chosen to be large in order to compare the overtones effect with the echoes at earlier times. }\label{fig:gravpotdef}
\end{figure*}
}

The pronounced sensitivity of overtone frequencies to small near-horizon deformations is not the only reason to study overtones. Another important aspect is the excitation factors of the overtones, which, as shown in particular examples in \cite{Silva:2024ffz}, may also increase significantly. Although the energy content of overtones is generally much smaller than that carried by the fundamental mode, this leads to a weaker observational prospect for overtones.

Now, suppose there is some form of deformation near the event horizon due to new physics. This leads to two effects: a modification of the signal in the early ring-down phase and a modification of the signal at very late times, when the signal is strongly damped. The latter effect is known as echoes \cite{Cardoso:2016rao} and has been widely studied in recent literature.
In this paper, we will demonstrate that the energy carried by overtones is much greater than that of echoes. Therefore, overtones offer a much more promising observational aspect of black hole spectra compared to echoes.

On fig.~\ref{fig:gravpotdef} we show the illustration of the same principle via the near-horizon deformation of the effective potential for gravitational perturbations of the Schwarzschild black hole by a Pöschl-Teller-like augmentation. The time-domain integration gives similar ringdown profiles for Schwarzschild and deformed potentials, but after the deduction of the fundamental-mode contribution via the Prony method one can see that the amplitude of the first overtone is many orders larger than the echoes produced at late times due to reflection from the Pöschl-Teller-like peak near the event horizon. Thus, the first overtone may be more perspective for probing the event horizon than the still elusive echoes. Another observational perspective of the overtones' outburst is related to the highly probable breakdown of isospectrality in a modified theory of gravity, like it happens for example in the Einstein-dilaton-Gauss-Bonnet and a number of other theories. In \cite{Jaramillo:2021tmt} it was shown that because of the disentanglement between axial and polar GW parities, the overtones outburst may already occur within the near-future detection range.

\section{Distinguishing near horizon deformations from astrophysical environment}

\begin{figure}
\resizebox{\linewidth}{!}{\includegraphics{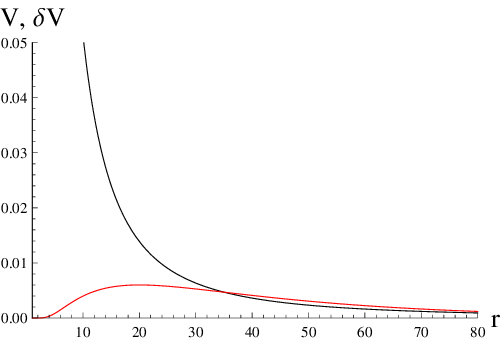}}
\caption{The effective potential for the $\ell=2$ gravitational perturbations of the Schwarzschild black hole (black) and the augmentation, defined by Eq.~\ref{augm} for $\delta=0.006$, $r_m=20$, $h=a=15$. The numerical values of the quasinormal modes are given in table~\ref{tabl:aug}.}\label{fig:augmentation}
\end{figure}

An important question remains whether it is possible that a tiny perturbation of the effective potential in the far zone, owing to black holes environment, such as an accretion disk, also produces the outburst of overtones \cite{Jaramillo:2021tmt,Jaramillo:2020tuu}? If so, then it would be difficult to distinguish the latter from the near-horizon deformations.
For this purpose as an example we consider a simple augmentation of the Schwarzschild potential in the form given by Eq.~(\ref{augm}).

\begin{table*}
\resizebox{\textwidth}{!}{
\begin{tabular}{c@{\hspace{1em}}c@{\hspace{1em}}c@{\hspace{1em}}c@{\hspace{1em}}c}
\hline
$n$ & Schwarzschild & $\delta=0.0006$, $r_m=20r_0$ & $\delta=0.006$, $r_m=20r_0$ & $\delta=0.006$, $r_m=50r_0$\\
\hline
$0$ & $0.747343 - 0.177925\imo$ & $0.747345 - 0.177928\imo$ & $0.747360 - 0.177957\imo$ & $0.747351 - 0.177937\imo$ \\
$1$ & $0.693422 - 0.547830\imo$ & $0.693457 - 0.547810\imo$ & $0.693776 - 0.547640\imo$ & $0.693555 - 0.547738\imo$ \\
$2$ & $0.602107 - 0.956554\imo$ & $0.602036 - 0.956371\imo$ & $0.601425 - 0.954672\imo$ & $0.601761 - 0.955865\imo$ \\
$3$ & $0.503010 - 1.410296\imo$ & $0.502438 - 1.410320\imo$ & $0.496956 - 1.410590\imo$ & $0.500803 - 1.410580\imo$ \\
$4$ & $0.415029 - 1.893690\imo$ & $0.414407 - 1.894740\imo$ & $0.409708 - 1.905080\imo$ & $0.413087 - 1.897930\imo$ \\
$5$ & $0.338599 - 2.391216\imo$ & $0.339163 - 2.393330\imo$ & $0.346480 - 2.409976\imo$ & $0.341543 - 2.398760\imo$ \\
\hline
\end{tabular}
}
\caption{Dominant modes of axial gravitational perturbations for the Schwarzschild black hole ($\ell=2$) compared with the ones for the potential deformed by augmentation (\ref{augm}) with $h=a=15$: the second column corresponds to the maximum deformation of the order of 0.1\% of the Schwarzschild potential peak and the last two columns correspond to higher deformation of the order of 1\% placed at different distances from the black hole. The modes are calculated using the Frobenius method in the units $r_0=2M=1$.}\label{tabl:aug}
\end{table*}

We take $\delta=0.006$, which is two orders smaller than the height of the main Schwarzschild peak for $\ell=2$ (see Fig.~\ref{fig:augmentation}). Note, that the maximum deformation is of the same order as the one considered in Table I of \cite{Jaramillo:2021tmt}.
Having in mind various types of possible astrophysical environment, choosing the value of $\delta \sim 10^{-3}$ is excessively big.
For example, the Shakura–Sunyaev disk model typically assumes that density of the accreting matter for the stellar-mass black holes, $\rho\leq100 g/cm^3$ (see \cite{Abramowicz:2011xu,Penna_2011} for review), which is more than 10 orders of magnitude smaller than the black-hole ``density''. For larger black holes the effects due to accreting matter are even smaller.
We have considered such big value of $\delta$ not only as a stress test for distinguishing the environmental effects, but also not to resort to time consuming computations of tiny changing of the frequencies.

Such astrophysically big deformation of the potential leads to very small corrections of the first several overtones (see table~\ref{tabl:aug}) and the corrections become even smaller if we shift the deformation farther from the black hole. Unlike the deformation considered in \cite{Jaramillo:2021tmt,Jaramillo:2020tuu}, the deformation (\ref{augm}) by construction does not affect the near-horizon behavior of the effective potential. Therefore we conclude that the phenomenon of the overtone outburst indeed happens due to deformations near the horizon.

If one takes sufficiently large values of $\delta$, providing the bump's height of the same order as the main peak, the overtones certainly become much more sensitive to such a modification, \eg for $\delta=0.1$ the real part of the higher overtones can be several times smaller than the Schwarzschild value. However, such a modification of the effective potential cannot be considered as a small deformation due to environment and is obviously physically irrelevant. These results are in agreement with the previous works: In~\cite{Cheung:2021bol} it was shown that for small bumps in the far zone, no outburst of overtones are observed, while in~\cite{Berti:2022xfj} the stability of the fundamental mode was found even for relatively large bumps.

\section{Conclusions}

We have shown that a small deformation of the near-horizon geometry of a black hole leads to a very strong change of overtones, while the fundamental mode remains almost unchanged. It immediately follows that first few overtones can probe the event horizon geometry which potentially could be seen at the earlier stage of quasinormal ringing \cite{Giesler:2019uxc}.

The observed here phenomenon of high sensitivity of lowest overtones to small deformations of the near horizon geometry is definitely connected with the so called ``overtones instability'' discussed in \cite{Jaramillo:2021tmt,Jaramillo:2020tuu}. There the effective potential was deformed by a sinusoidal function, so that this would correspond to the small deformation of the geometry not only near the event horizon, but in the whole space.
Moreover, in \cite{Jaramillo:2021tmt,Jaramillo:2020tuu} the derivative of the deformation function does not vanish at the event horizon, implying nonsmall changes of the near-horizon geometry, and the impact of the deformation obviously increases when the ``high-frequency'' deformation is considered.

Such a setup does not allow one to understand which kind of deformations produce the outburst of overtones, which is the main question of our consideration. Here we have shown that very small deformations in a relatively small region near the event horizon are sufficient for such an outburst of overtones, while the deformations at a distance from the black hole need to be very large and physically irrelevant in order to produce a similar effect. Having a number of overtones in addition to the fundamental mode and using the strict hierarchy of coefficients of the parametrization \cite{Rezzolla:2014mua,Konoplya:2016jvv}, one could, in principle, constrain the allowed black-hole geometry.

The high sensitivity of the overtones due to the small nonsmooth deformations, which is also the case of the random deformation probes of \cite{Jaramillo:2020tuu}, were studied also in earlier works \cite{Berry:1982,Nollert:1996rf,Aguirregabiria:1996zy}.
The crucial difference between our approach and the above works is that we considered smooth deformations distributed over some region near the horizon (when considering the horizon contribution, see Figs.~\ref{fig:scalarpot},~\ref{fig:gravpotdef}) and in the far zone (when considering the astrophysical environment, Eq.~\ref{augm}). This must be more appropriate to real astrophysical distribution of gravitational and matter fields than the nonlocalized deformations of \cite{Jaramillo:2021tmt,Jaramillo:2020tuu} or nonsmooth highly localised (``ultraviolet'') deformations considered in \cite{Jaramillo:2020tuu,Berry:1982,Nollert:1996rf,Aguirregabiria:1996zy}. After all, the latter high-frequency deformations of the black-hole spacetime would require high energies which could hardly be ascribed to an astrophysical environment in the linear regime.

It's also crucial to recognize that the influence of the near-horizon geometry, reflecting in the initial overtones, isn't restricted solely to minor deviations from the Schwarzschild/Kerr geometries in the near-horizon region. This is because strong constraints on the near-horizon geometry of black holes remain elusive. Since the initial version of this work was published on arXiv, several specific instances highlighting the sensitivity of overtones have been examined.
An outburst of overtones is expected for virtually any metric that is modified in the near-horizon zone. The only difference lies in the rate at which the deviation occurs, meaning the overtone number at which the deviation becomes significant. From a practical perspective, if such a strong deviation does not appear within the first few overtones, it will not have any potential for observational relevance. Black hole metrics in various alternative theories of gravity often differ significantly in the near-horizon zone, indicating a promising future direction for studying overtones within alternative gravity theories.

\begin{table}
\begin{tabular}{|c|c|c|}
  \hline
  \hline
  Metric & $n$ & publication \\
   \hline
   \hline
  Hayward  & 1 & \cite{Konoplya:2022hll,Konoplya:2023ppx}\\
  Bonnano-Reuter & 1 & \cite{Konoplya:2022hll} \\
  Dymnikova & 3 & \cite{Konoplya:2023aph} \\
  Bardeen & 4 & \cite{Konoplya:2023ahd,Bolokhov:2023ruj} \\
  Einstein-Weyl & 4 & \cite{Konoplya:2022iyn} \\
  D-dimensional Einstein-(GB)-AdS & 1 & \cite{Konoplya:2023kem} \\
  sub-Planckian curvature & 4 & \cite{Zhang:2024nny} \\
  effective quantum gravity & 2 & \cite{Konoplya:2024lch} \\
  quantum Oppenheimer-Snyder model & 2 & \cite{Zinhailo:2024kbq} \\
  \hline
  \hline
\end{tabular}
\caption{Outburst of overtones for various theories of gravity and the minimal number of $n$ which shows distinctive behavior.}\label{tableResults}
\end{table}

In cases where deformations were significant in the near-horizon area but minor or moderate in the radiation zone, notable behaviors, even in the first and second overtones, were observed. Thus, for the Hayward, Bonanno-Reuter, and Schwarzschild-AdS black holes the qualitatively new behavior occurs already at $n=1$ where a purely imaginary modes may appear, while higher $n$ differ from their Schwarzschild values by tens and, sometimes, more than $100\%$ percents. We can consolidate these specific instances across various alternative theories of gravity in Table \ref{tableResults}. However, the majority of the aforementioned works, while capturing the qualitative behavior of the overtone outburst, focus on perturbations of test fields. Thus, this program is in the very beginning, because no thorough investigation of overtones' behavior for gravitational perturbations have been made in alternative theories of gravity.

Note that, while overtones are not the only theoretical method for probing near-horizon geometry, they may be one of the most promising approaches. This is because echoes correspond to a much weaker signal, while the black hole shadow primarily provides information about the geometry of the photon sphere. However, in the case of near-extremal rotating black holes, the photon sphere approaches the event horizon. In this regime, the shadows cast by black holes, as well as the fundamental mode, may indeed reveal information about the near-horizon geometry \cite{Horowitz:2023xyl,EventHorizonTelescope:2019pgp}.

\acknowledgments
A.~Z. was supported by Conselho Nacional de Desenvolvimento Científico e Tecnológico (CNPq).

\bibliographystyle{unsrt}
\bibliography{bibliography}

\end{document}